\documentclass[%
 aip,
 jcp,
 amsmath,amssymb,
 reprint,%
longbibliography,
]{revtex4-1}

\usepackage{graphicx}
\usepackage{dcolumn}
\usepackage{bm}
\usepackage[mathlines]{lineno}

\usepackage[utf8]{inputenc}
\usepackage[T1]{fontenc}
\usepackage{mathptmx}
\usepackage{etoolbox}
\makeatletter
\def\@email#1#2{%
 \endgroup
 \patchcmd{\titleblock@produce}
  {\frontmatter@RRAPformat}
  {\frontmatter@RRAPformat{\produce@RRAP{*#1\href{mailto:#2}{#2}}}\frontmatter@RRAPformat}
  {}{}
}%

\usepackage{hyperref}

\makeatother
\begin{document}

\preprint{AIP/123-QED}

\title{Ionomer structure and component transport in the cathode catalyst layer of PEM fuel cells: A molecular dynamics study}
\author{Yichao Huang}
\affiliation{State Key Laboratory of Engines, Tianjin University, Tianjin, 300350, China.}
\author{Panagiotis E. Theodorakis}
\affiliation{Institute of Physics, Polish Academy of Sciences, Al.~Lotnik\'{o}w 32/46, 02-668 Warsaw, Poland}
\author{Zhen Zeng}
\affiliation{State Key Laboratory of Engines, Tianjin University, Tianjin, 300350, China.}
\affiliation{National Industry-Education Platform of Energy Storage, Tianjin University, Tianjin 300350, China}
\author{Tianyou Wang}
\affiliation{State Key Laboratory of Engines, Tianjin University, Tianjin, 300350, China.}
\affiliation{National Industry-Education Platform of Energy Storage, Tianjin University, Tianjin 300350, China}
\author{Zhizhao Che}
\email{chezhizhao@tju.edu.cn}
\affiliation{State Key Laboratory of Engines, Tianjin University, Tianjin, 300350, China.}
\affiliation{National Industry-Education Platform of Energy Storage, Tianjin University, Tianjin 300350, China}
\date{\today}

\begin{abstract}
The transport of water and protons in the cathode catalyst layer (CCL) of proton exchange membrane (PEM) fuel cells is critical for cell performance, but the underlying mechanism is still unclear. Herein, the ionomer structure and the distribution/transport characteristics of water and protons in CCLs are investigated via all-atom molecular dynamics simulations. The results show that at low water contents, isolated water clusters form in ionomer pores, while proton transport is mainly via the charged sites of the ionomer side chains and the Grotthuss mechanism. Moreover, with increasing water content, water clusters are interconnected to form continuous water channels, which provide effective paths for proton transfer via the vehicular and Grotthuss mechanisms. Increasing the ionomer mass content can enhance the dense arrangement of the ionomer, which in turn increases the density of charge sites and improves the proton transport efficiency. When the ionomer mass content is high, the clustering effect reduces the space for water diffusion, increases the proton transport path, and finally decreases the proton transport efficiency. By providing physics insights into the proton transport mechanism, this study is helpful for the structural design and performance improvement of CCLs of PEM fuel cells.
\end{abstract}

\maketitle

\section{Introduction}\label{sec:1}
As a type of green and clean energy conversion device, proton exchange membrane (PEM) fuel cell has received widespread attention due to its high efficiency and broad application prospects \cite{ogungbemi21, stambouli11,wang11, wang20}. A PEMFC consists of PEM, catalyst layers (CL), gas diffusion layers (GDL), and bipolar plates (BP) \cite{jiao21}. Among those, the CL, as the place where electrochemical reactions take place, plays a critical role in improving the efficiency of fuel cells. It is mainly composed of carbon-supported platinum (Pt/C) and ionomer covering platinum \cite{offer09, soboleva11}. Ionomers, which have charged sites on their side chains, allow protons to travel through the catalyst layer. Water molecules act as binders to fill the pores of the ionomers to provide proton transport channels. The amount of water is crucial for the ionomer. Too little water will not be able to fully infiltrate the ionomer, increasing the proton transport resistance; too much water will cause water flooding of the cathode catalyst layer (CCL) and reduce the fuel cell performance \cite{deng20}. Therefore, the internal structure and the transport process in CCLs are important factors affecting fuel cell performance \cite{deng20, holdcroft13, kusoglu17, lyulin20, so19, sun13}.

The transport process in CCL includes proton transport, electron conduction, water diffusion, and oxygen transport. The ionomer microstructure in CCLs will affect the transport of protons. A low ionomer content will make too little ionomer contact and cannot completely surround the platinum particles, resulting in a decrease in the electric conductivity \cite{xue22}, while too much ionomer will lead to agglomeration, which will also reduce the coverage of the catalyst and lead to a decrease in the electric conductivity \cite{andersen16}. The ionomer/catalyst ratio and ionomer thickness, which correspond to the ionomer content, will affect the transfer of components inside the ionomer \cite{lee22, wang21md}. Therefore, the content of ionomers is an important factor for the transport characteristics in CLs \cite{wang19}. Via experiments and numerical simulations, the effects of ionomer content on conductivity and cell performance have been evaluated at macroscales \cite{kim10, li03, suzuki11, wuttikid17, xing14}. However, since it is difficult to directly observe the characteristics of its nanostructures and mass transfer through experiments, molecular dynamics (MD) methods are often employed to investigate the mechanisms of mass transfer \cite{atrazhev17, lamas06, sengupta18}. Through MD simulations, it has been found that the proton transport path and transport mechanism have close correlations with the water content. Increasing the water content will make the water change from isolated clusters to continuous water channels, which will affect the transport of protons in ionomers \cite{fan19water}. Excessive water will fill the free volume between ionomers, increasing the transfer resistance \cite{kurihara17}. Due to the hydrophilicity of platinum and the hydrophobicity of carbon, the substrate of carbon-supported platinum also affects the diffusion of water molecules, and then affects the proton transport. In CLs, there are mainly two forms of proton transport, namely, the vehicular and Grotthuss mechanisms \cite{jiao11}. In particular, protons are transported by the vehicular mechanism through the combination with different forms of water clusters as ${{\text{H}}_{3}}{{\text{O}}}^{+}$, ${{\text{H}}_{5}}{{\text{O}}_{2}}^{+}$, and ${{\text{H}}_{9}}{{\text{O}}_{4}}^{+}$ \cite{pivovar06}. With the increase in the water content, continuous water channels will form and connect the charged sites on ionomer side chains. In this condition, protons will be transported by hopping among water molecules, namely, the Grotthuss or hopping mechanism \cite{agmon1995}. Kang et al. found that the dense water structure on the platinum surface promotes the Grotthuss mechanism of protons \cite{kang20}.

Previous studies have shown that understanding the effects of ionomer content and water content in CCLs on the transport of water and protons is of great significance for analyzing the component transport mechanism and improving the transport efficiency. In this study, the nanostructure of hydrated ionomer on carbon-supported platinum substrates in CCLs is studied by all-atom MD simulations. The influence of ionomer content and water content on the micro-/nano-structure of ionomers are unveiled by analyzing the radial distribution function (RDF), density distribution, and coordination number (CN). Finally, the transport characteristics of water and proton in the ionomer are investigated by calculating the self-diffusion coefficients including two mechanisms of proton transport (i.e., Grotthuss and vehicular mechanisms). The morphological characteristics of water clusters and the corresponding effects on the proton transport mechanism are also studied by analyzing the number of hydrogen bonds between hydronium ions and water or sulfonic acid groups.

\section{MD simulation methods and details}\label{sec:2}
\subsection{System configuration}\label{sec:2.1}
\begin{figure}
  \centering
  \includegraphics[scale=0.6]{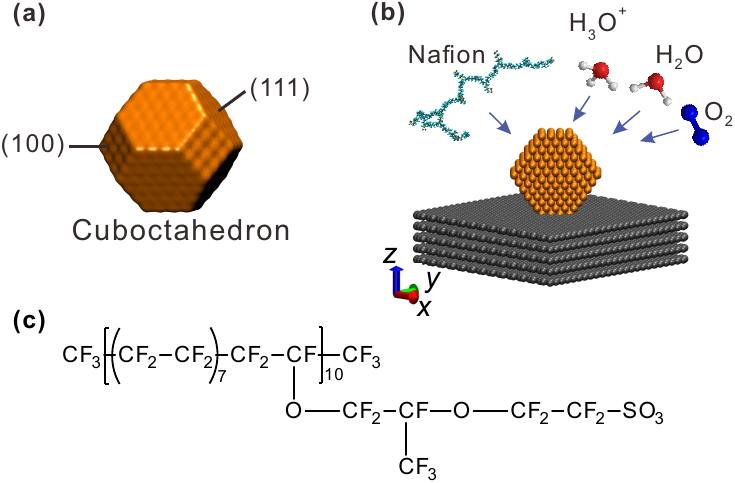}
  \caption{(a) Structure of the cuboctahedron platinum particle; (b) Schematics of Pt/C and ionomer in CLs; (c) Chemical structure of PFSA ionomer.}\label{fig:01}
\end{figure}

The system consists of a carbon substrate, a platinum (Pt) particle, PFSA (perfluorosulfonic acid, Nafion) ionomer, water, hydronium (${{\text{H}}_{3}}{{\text{O}}}^{+}$), and oxygen molecules, as shown in Figure \ref{fig:01}b. The carbon substrate is made up of five layers of carbon atoms, each with a separation of 3.354 {\AA} between them. Each carbon layer contains 1344 carbon atoms, which were set in the $x$-$y$ plane with the size of 58 {\AA} $\times$ 59 {\AA}. The platinum particle has the shape of a truncated octahedron, including eight (111) planes and six (100) planes, as shown in Figure \ref{fig:01}(a). Former studies have pointed out that this platinum particle shape is the most consistent with the platinum particles in actual PEM fuel cells \cite{cheng10}. The Pt particle consists of 586 platinum atoms, approximately 2.35 nm in diameter.
The Pt particle was placed on top of the carbon substrate, which was placed at the bottom of the simulation box. Periodic boundary conditions are applied in all three Cartesian directions of the simulations box (Figure \ref{fig:01}).

To simulate the ionomer electrolyte with different mass contents, different numbers of PFSA chains were placed on the Pt/C substrate. Each PFSA chain contains 10 repeating units and each unit contains a sulfonic acid group ($\text{SO}_{3}^{-}$), as shown in Figure \ref{fig:01}(c). The equivalent weight (EW) of PFSA ionomer is 1144 g/mol. The number of PFSA ionomer chains was varied from 4--16, corresponding to the ionomer mass content ($\varphi$) in the total solid mass of CCL from 19.1--48.5\%. The ionomer mass content ($\varphi$) is defined as
\begin{equation}\label{eq:phi}
  \varphi =\frac{{{m}_{\text{ionomer}}}}{ {{m}_{\text{ionomer}}}+{{m}_{\text{C-plate}}}+{{m}_{\text{Pt-particle}}}},
\end{equation}
where ${{m}_{\text{ionomer}}}$, ${{m}_{\text{C-plate}}}$, and ${{m}_{\text{Pt-particle}}}$ are the masses of ionomer, carbon plate, and platinum particle, respectively. To maintain the system's electrical neutrality, hydronium ions (${{\text{H}}_{3}}{{\text{O}}}^{+}$) were added to the system according to the ionomer mass contents (i.e., 10 ${{\text{H}}_{3}}{{\text{O}}}^{+}$ for each PFSA chain). The water content ($\lambda$) is defined as the total number of water molecules and hydronium ions per sulfonate group:
\begin{equation}\label{eq:lambda}
  \lambda =\frac{ {{N}_{{{\text{H}}_{ {2}}}\text{O}}}+{{N}_{{{\text{H}}_{ {3}}}{{\text{O}}^{ {+}}}}}}{{{N}_{\text{SO}_{3}^{-}}}},
\end{equation}
where ${{N}_{{{\text{H}}_{ {2}}}\text{O}}}$, ${{N}_{{{\text{H}}_{ {3}}}{{\text{O}}^{ {+}}}}}$, and ${{N}_{\text{SO}_{3}^{-}}}$ are the number of water molecules, hydronium ions, and sulfonate groups. The water content was varied from $\lambda$ = 3 to 20 in this study, while the number of oxygen molecules ${{N}_{{{\text{O}}_{ {2}}}}}$ was fixed at 80.

\subsection{Simulation details}\label{sec:2.2}
To describe the intra- and inter-molecular interactions in the system (except water, hydronium ion, and platinum), we employed a modified DREIDING force field \cite{mayo90}, which has been used to describe ionomer electrolyte systems in PEM fuel cells successfully \cite{wang21md, kang20, fan20, kwon19, mabuchi14}. Water molecules and platinum particles are described by the F3C and EAM force field \cite{levitt97}. The classical hydronium model is applied to describe the hydronium ion \cite{jang04}.

The simulations were performed using LAMMPS package \cite{plimpton95}. The velocity-Verlet algorithm \cite{Swope82} was used to combine the atomic motion equations with a time step of 1 fs. The particle--particle particle--mesh (PPPM) method with an accuracy of 0.0001 was used to calculate the long-range electrostatic interactions. The temperature and pressure of the system were controlled using the Nos\'{e}--Hoover thermostat and barostat with relaxation times of 0.1 and 1 ps, respectively.

After initializing the molecules (except water and oxygen molecules) into the simulation box, annealing procedures were performed to reliably obtain their equilibrium structures. The annealing procedures are as follows: (a) Starting from the initial structure, the system temperature was increased from 0 K to 353 K in 300 ps. (b) The system temperature was increased from 353 K to 1000 K in 300 ps. (c) NVT simulation was performed to maintain the temperature at 1000 K for 300 ps. (d) The system temperature was decreased from 1000 K to 353 K in 300 ps. (e) NVT simulation was performed to maintain the temperature at 353 K for 300 ps. (f) Steps (b-e) were repeated for another two times. (The temperature cycling can better spread the ionomer on the substrate, creating proper initial conditions for obtaining the subsequent equilibrium structure). (g) The ionomer structure was equilibrated over 600 ps of NVT simulation at 353 K.

After the annealing procedures of the PFSA ionomer structure, water and oxygen molecules were added into the system for the subsequent simulations. An NVT MD simulation of 1 ns was performed to relax water and oxygen molecules preliminarily. Then, MD simulation of 5 ns was performed under an isothermal--isobaric ensemble (NPT) with the pressure in the $z$ direction at 1 bar and the temperature at 353 K. The above steps ensure that the whole system will reach its equilibrium. After the equilibrium procedures, an NVT MD simulation of 15 ns was performed for data sampling, and the density-related data was collected within the last 5 ns. These steps in our simulations can allow us to better study the transport characteristics of water and hydronium ions in CCLs and the effects of water contents on ionomer structures.

\section{Results and discussion}\label{sec:3}
\subsection{Equilibrium structure and density}\label{sec:3.1}
\begin{figure}
  \centering
  \includegraphics[scale=0.57]{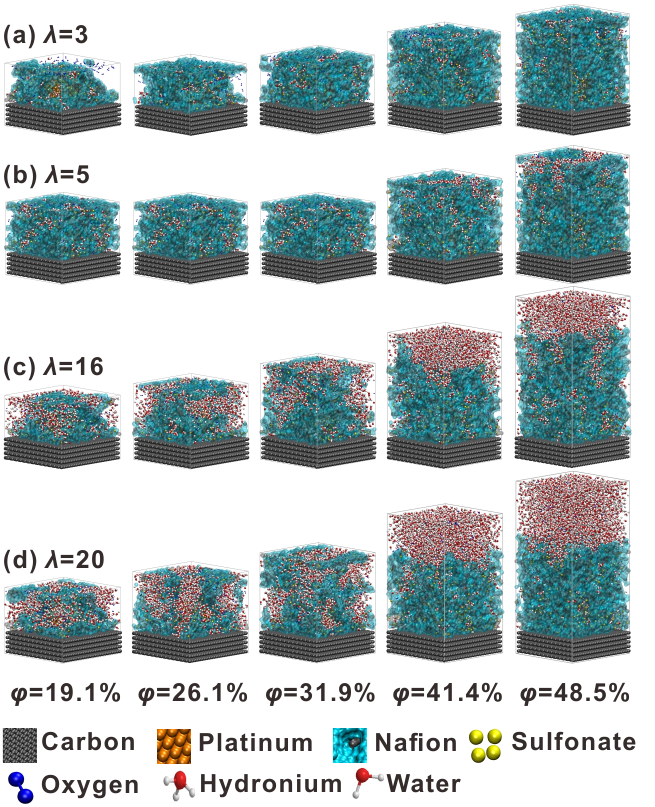}
  \caption{Snapshots of equilibrium systems with different levels of water content ($\lambda$) and ionomer mass content ($\varphi$): (a) $\lambda$ = 3; (b) $\lambda$ = 5; (c) $\lambda$ = 16; (d) $\lambda$ = 20.}\label{fig:02}
\end{figure}

Typical snapshots of the CLs at the end of the MD simulations are shown in Figure \ref{fig:02}. We can see that the PFSA ionomer is well spread on the Pt/C substrate, water and oxygen molecules diffuse freely in the pores formed by the ionomer spreading, and at the upper surface of the ionomer layer, some water molecules aggregate. Under the same level of water content ($\lambda$), as the ionomer mass content ($\varphi$) increases, the thickness of the ionomer layer becomes larger. With the dense stacking of the ionomer molecules, the thickness and the number of ionomer molecules show a positive correlation. Under the same ionomer mass content ($\varphi$), increasing the water content will lead to the formation of hydrophilic and hydrophobic regions. Isolated water clusters are connected to form a continuous water phase which fills the free space formed by ionomer stacking. Moreover, the high water content will also affect the spreading of the ionomer, further changing the ionomer thickness.

\begin{figure}
  \centering
  \includegraphics[scale=0.8]{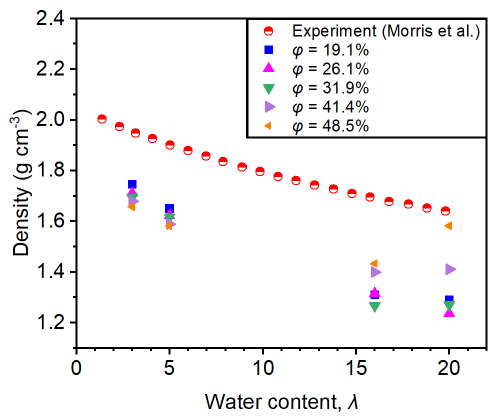}
  \caption{Densities of hydrated ionomer compared with experimental results reported by Morris et al. \cite{morris93}.}\label{fig:03}
\end{figure}

The ionomer density is a key parameter to quantify the equilibrium state of the ionomer. The ionomer density was obtained by calculating the mass of the hydrated ionomers (including ionomer, water, and hydronium ions) divided by their total volume. As shown in Figure \ref{fig:03}, the ionomer density is mainly affected by the water content. At low ionomer mass content, increasing the water content will hydrate the ionomer sufficiently, resulting in an increase in the volume of the ionomer and thus a decrease in the density, which is consistent with the experimental results reported by Morris et al.\ \cite{morris93}. With the increase in the ionomer mass content, the influence of the water content on the ionomer density will be gradually reduced, because the dense stack of ionomer will have an adverse effect on the swelling caused by hydration. Therefore, the density decline caused by the increased water content will become smaller with the increase in the ionomer mass content. When the ionomer mass content is too high ($\varphi>41.4\%$), the ionomer will produce the clustering effect, which will hinder the transport of components in the ionomer layer. The swelling caused by hydration will be further weakened, and the water clusters gathered in the free space above the ionomer will further reduce the ionomer volume, resulting in a large increase in the ionomer density.

\subsection{Morphology and distribution of ionomer}\label{sec:3.2}
\begin{figure*}
  \centering
  \includegraphics[width=1.8\columnwidth]{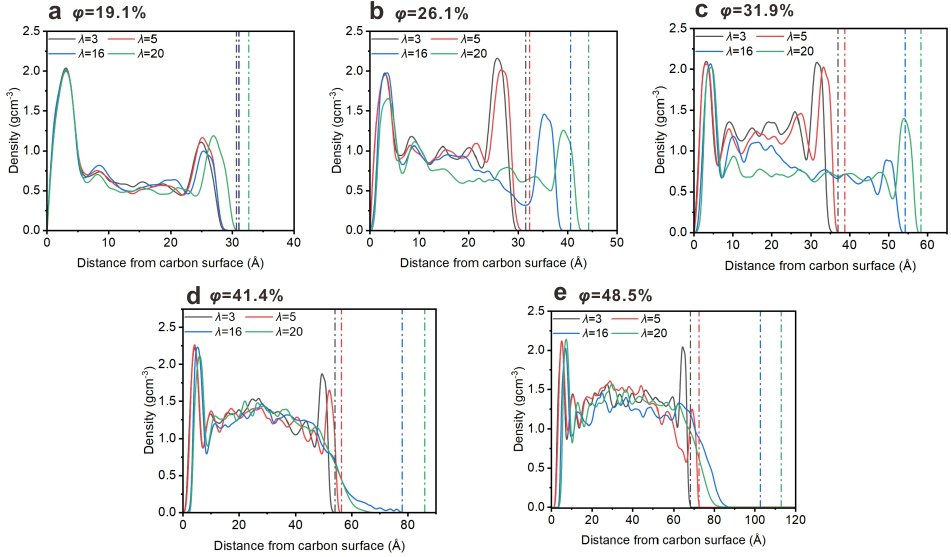}
  \caption{Density distributions of hydrated ionomer on the Pt/C substrate along the thickness (z) direction for different ionomer mass contents: (a) 19.1\%, (b) 26.1\%, (c) 31.9\%, (d) 41.4\%, (e) 48.5\%. The dashed-dotted lines indicate the position of the upper boundary for different ionomer mass contents, and the line colors are consistent with the colors in the legend.}
  \label{fig:04}
\end{figure*}

To investigate the effect of the CL composition on the morphology and distribution of ionomer, the water content ($\lambda$) and the ionomer mass content ($\varphi$) were changed, and the last 5 ns of the sampling stage was used to obtain density-related parameters. Sampling was performed every 50 ps, and 100 configurations were averaged to obtain the density distribution of ionomer in the $z$ direction from the carbon surface. The density distributions of the ionomer under different water contents and ionomer mass contents are shown in Figure \ref{fig:04}. It can be found that an ultrathin ionomer layer ($\sim$ 0.75 nm) is formed on the carbon substrate, which is consistent with the results reported in Ref.~\citenum{fan20}. A dense ultrathin ionomer film ($\sim$ 0.75 nm) is also formed close to the upper boundary, because of the interaction between the ionomer skeleton and the upper boundary carbon. These two ionomer films result in two peaks in the density distribution curves of hydrated ionomer. The first peak (close to the lower boundary, $z = 0$) is slightly higher than the second peak (close to the upper boundary). This is because at the lower side of the platinum particle, the ionomer carbon skeleton is affected by the carbon substrate, and the side chain is affected by the catalyst, hence the two types of interactions make the ionomer denser. With increasing the ionomer mass content, the second ultra-thin ionomer layer shifts in the $z$ direction. And because the position of the second peak is further away from the platinum particle, the interaction between the ionomer side chain and platinum is weakened, thus decreasing the second peak height. With increasing the water content, the formation of water channels and the stratification between the water phase and ionomer phase are enhanced. Water molecules will fill into the ionomer layer, increase the ionomer thickness, and also affect the position of the second peak. Water molecules will also aggregate on the upper surface of the ionomer layer, hindering the interaction between the ionomer skeleton and the upper carbon. The interaction of platinum with the ionomer side chain is hindered by the increase of ionomer layer thickness, and the interaction of the ionomer skeleton with the upper carbon is hindered by the water molecules. These two factors lead to the disappearance of the thin ionomer film close to the upper boundary. The distribution of the ionomer tends to be homogeneous and the ionomer layer becomes thicker.

\begin{figure*}
  \centering
  \includegraphics[scale=1]{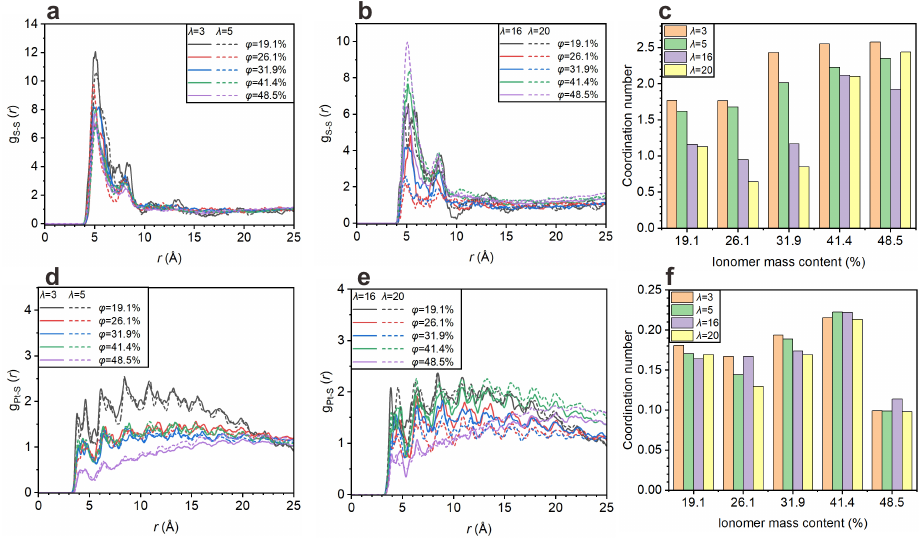}
  \caption{(a,b) RDFs of S--S pairs at different ionomer mass contents. (c) CN of S--S. (d,e) RDFs of Pt--S pairs at different ionomer mass contents. (f) CN of Pt--S. (a) and (d) show the results at low water contents ($\lambda$ = 3, and 5), and (b) and (e) show the results at high water contents ($\lambda$ = 16 and 20).}\label{fig:05}
\end{figure*}

The structural and distribution characteristics of ionomer are further explored by the radial distribution function (RDF), which calculates the probability of finding atoms B around atom A with a separation distance of $r$ in the equilibrium structure. The RDF is calculated as
\begin{equation}\label{eq:01}
  {{g}_\text{A-B}}(r)={\left( \frac{{{n}_\text{B}}}{4\pi {{r}^{2}}dr} \right)}/{\left( \frac{{{N}_\text{B}}}{V} \right)}
\end{equation}
where $n_\text{B}$ is the number of atoms B located at a distance $r$ in a shell of thickness $dr$, $N_\text{B}$ is the number of atoms B in the system, and $V$ is the total volume of the system. Since the water content and the ionomer mass content affect the structure and distribution of ionomer, we analyze the RDF of sulfur (in the side chain of PFSA ionomer) and sulfur, sulfur and platinum, i.e., $g_\text{s-s}$ and $g_\text{Pt-s}$. Figure \ref{fig:05}(a-b) shows the sulfur--sulfur radial distribution function $g_\text{s-s}$ under different water contents and ionomer mass contents. The coordination number (CN) was also obtained by analyzing the RDF to get the position of the first solvation shell, and then integrating the RDF. The cutoff distance was set at 6.85 {\AA} to calculate the CN of S--S pairs by integrating the RDF, as shown in Figure \ref{fig:05}(c). It can be found that CN shows an increasing trend with increasing the ionomer mass content, revealing that the cross-linking effect of S--S is enhanced, the space between sulfonic acid groups becomes smaller, and the charged sites become closer. Therefore, protons can transport from one water cluster to another through the attraction of charged sites, which is conducive to the Grotthuss mechanism. With the increase in the water content $\lambda$, CN decreases, indicating that the gradual formation of water channels will hinder the aggregation of sulfonic acid groups, resulting in further separation of the water phase and ionomer phase, reducing the role of the Grotthuss mechanism in proton transport.

To further illustrate that the Grotthuss mechanism is affected by the distribution of sulfonic acid groups, the RDF of sulfur and platinum ($g_\text{Pt-s}$) is analyzed, see Figure \ref{fig:05}(d-e). It can be found that the variation of the water content has little effect on the aggregation of sulfonic acid groups on the surface of platinum particles (see the difference between the solid lines and dashed lines in Figure \ref{fig:05}(d-e)), while the change of ionomer mass content plays a leading role (see the difference between different line colors in Figure \ref{fig:05}(d-e)). With increasing the ionomer mass content, the ionomer accumulates densely on the Pt/C substrate, and the probability of finding sulfonic acid groups around the platinum particle increases, which further facilitates protons to reach the catalyst surface through the Grotthuss mechanism. However, when the ionomer mass content reaches 48.5\%, the CN value shows a sharp decrease. This is because, at the ionomer mass content of 41.4\%, the spreading of ionomer on the Pt/C substrate leads to a dense distribution. With further increasing the ionomer mass content, the S--S cross-linking also increases, which changes the structure of the originally well-spreading ionomer film. The ionomer film curls internally and develops into a cluster (referred to as clustering effect hereafter), which reduces the surface area of the ionomer for external interaction, enhances the compactness of the ionomer film, and finally, hinders the transport of water and protons. Therefore, a high ionomer mass content results in the separation between the ionomer and the substrate, and the CN decreases correspondingly.

\begin{figure}
  \centering
  \includegraphics[scale=0.45]{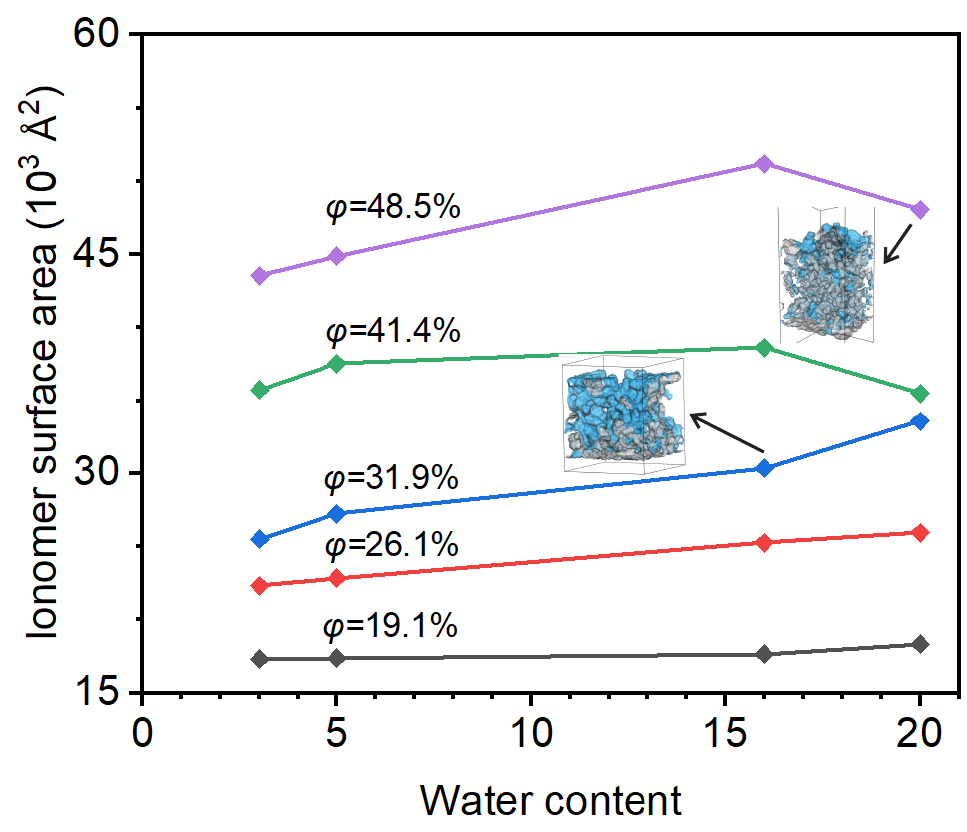}
  \caption{Variation of ionomer surface against water content ($\lambda$) area at different ionomer mass content ($\varphi$).}\label{fig:06}
\end{figure}

To further demonstrate that the water content ($\lambda$) and ionomer mass content ($\varphi$) affect the stacking of the ionomer, the surface area of the ionomer was analyzed, as shown in Figure \ref{fig:06}. The surface area was obtained via the Connolly surface analysis \cite{connolly83} by setting a probe radius of 0.14 nm that represents the radius of ionomer. As shown in Figure \ref{fig:07}, the surface area of the ionomer is mainly influenced by the ionomer mass content $\varphi$, while the effect of the water content $\lambda$ is weak. When $\varphi$ is less than 40\%, the surface area of the ionomer increases slightly with the increase in the water content. Increasing the ionomer surface area will increase the reaction area, which is beneficial to the encapsulation of water clusters to ionomer side chains and the transport of protons. The snapshot of the well-stacked hydrated ionomer in the condition of $\varphi=31.9\%$ and $\lambda=16$ is also shown in Figure \ref{fig:06}. When $\varphi$ is larger than 40\%, the previously mentioned clustering effect appears, and the high water content will further affect the surface area of the ionomer. As the clustering effect is intensified, the surface area of the ionomer decreases sharply, which is not conducive to proton transport.

\subsection{Water transport and morphology properties}\label{sec:3.3}
\begin{figure*}
  \centering
  \includegraphics[width=1.5\columnwidth]{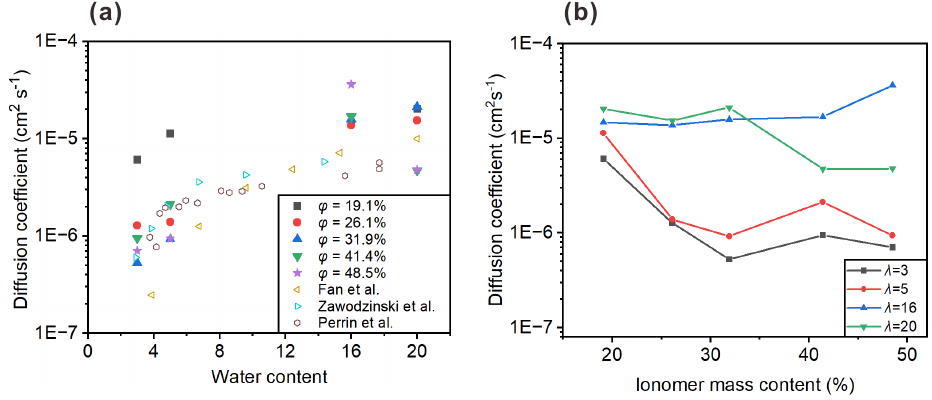}
  \caption{(a) Diffusion coefficient of water at different ionomer mass contents ($\varphi$) and water contents ($\lambda$). The results obtained in previous experimental and simulated results by Fan et al. \cite{fan19}, Zawodzinski et al. \cite{zawodzinski91}, and Perrin et al. \cite{perrin07} are also shown. (b) Diffusion coefficient of water against the ionomer mass content ($\varphi$).}\label{fig:07}
\end{figure*}

The distribution and transport of water and hydronium ions depends on the formation of water channels in the ionomer, which is affected by the water content and the ionomer mass content. In this study, the diffusion coefficient of water and hydronium ions was obtained according to
\begin{equation}\label{eq:02}
  D=\frac{1}{6N}\underset{t\to \infty }{\mathop{\lim }}\frac{1}{\Delta t}\left[ \sum\limits_{i=1}^{N}{{{(\overrightarrow{{{r}_{i}}}(t+\Delta t)-\overrightarrow{{{r}_{i}}}(t))}^{2}}} \right]
\end{equation}
where $\overrightarrow{{{r}_{i}}}$ is the position vector of target particles, $\overrightarrow{{{r}_{i}}}(t+\Delta t)-\overrightarrow{{{r}_{i}}}(t)$ refers to the displacement of target particles within $\Delta t$ time length, and $N$ means the total number of diffusion particles. The calculation of the mean square displacement parameters of water and hydronium ions starts from the sampling step and lasts for 10 ns. The calculated diffusion coefficient of water is also compared with the results from previous simulations and experiments \cite{fan19, perrin07, zawodzinski91}. With increasing water content, the diffusion coefficient of water molecules increases, which is consistent with previous studies \cite{fan19, perrin07, zawodzinski91}, as shown in Figure \ref{fig:07}. This is because increasing the water content will lead to the formation of water channels with stronger connectivity and larger volume, which is conducive to the water diffusion. When the ionomer mass content is between 26.1\% and 48.5\%, the water content level has a strong correlation with the water diffusion coefficient. When the ionomer mass content is low, the diffusion of water molecules is not limited by the ionomer layer because of the larger free volume (i.e., the space that is un-occupied by the ionomer and allows water molecules to diffuse through), and it is more conducive to water diffusion, as shown in Figure \ref{fig:07}(b). With the gradual increase of ionomer mass content, the ionomer is densely spread on the Pt/C substrate, and the low water content has little effect on the spread of the ionomer. Therefore, the water diffusion coefficient decreases due to the reduction in the free volume. In water-rich conditions, water molecules affect the spreading of ionomer, forming stratification of the ionomer phase and water phase. Therefore, the ionomer phase has a significant increase in thickness, so the water diffusion coefficient increases due to the increase in free volume.

\begin{figure}
  \centering
  \includegraphics[scale=0.5]{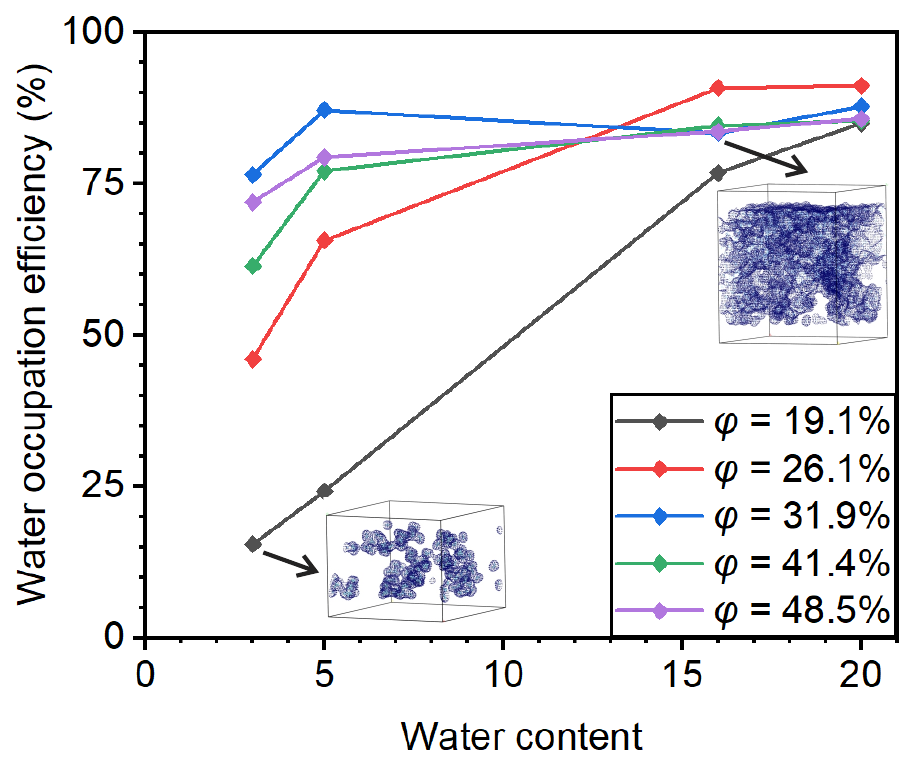}
  \caption{Variation of water occupation efficiency against the water content ($\lambda$) in different ionomer mass contents ($\varphi$).}\label{fig:08}
\end{figure}

To further explore the distribution of water molecules in the ionomer layer, we analyze the water occupation efficiency ($\varepsilon$), which is defined as
\begin{equation}\label{eq:03}
  \varepsilon =\frac{{{V}_{\text{water}}}}{{{V}_{\text{sys}}}-{{V}_{\text{ionomer}}}-{{V}_{\text{substrate}}}-{{V}_{\text{oxygen}}}}
\end{equation}
where ${{V}_{\text{sys}}}$, ${{V}_{\text{water}}}$, ${{V}_{\text{ionomer}}}$, ${{V}_{\mathrm{substrate}}}$, and ${{V}_{\text{oxygen}}}$ are the volumes of the system, water, ionomer, substrate, and oxygen, respectively. Except ${{V}_{\text{sys}}}$ using the volume of the simulation box, the other volumes are calculated via the Connolly volume analysis \cite{connolly83} with a probe radius of 0.14 nm. With increasing the water content, the water occupation efficiency gradually increases, as shown in Figure \ref{fig:08}. It is also significantly affected by the ionomer mass content. At a low water content, the water occupation efficiency at $\varphi$ = 19.1\% and 26.1\% is very low, indicating that the entire simulation system contains a large free space, so the water molecules diffuse relatively easier in this condition. By increasing the water content, the water occupation efficiency has a significant increment. When $\varphi$ reaches 31.9\%, the water occupation efficiency maintains at a high level at all water contents, because the ionomer spreads well, and the free space mainly exists in the ionomer pores. Hence, this ionomer mass content is more conducive to proton transport. When $\varphi$ is further increased, due to the clustering effect, the volume of pores in the ionomer is compressed, and the free space is further reduced. In this condition, the efficiency is mainly influenced by the ionomer mass content. If the water content comes to a certain value (i.e., $\lambda = 5$ as shown in Figure \ref{fig:08}), it is enough to fill the free space that water molecules can reach, so the subsequent increase of water content has little effect on the water occupation efficiency.

\begin{figure*}
  \centering
  \includegraphics[width=1.8\columnwidth]{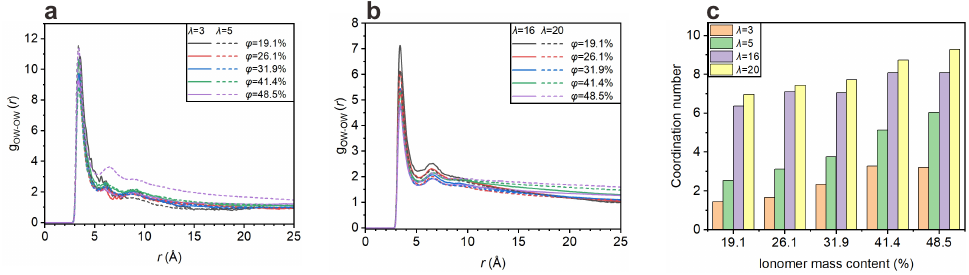}
  \caption{(a,b) RDFs of OW--OW pairs at different ionomer mass contents at low and high water contents: (a) low water content $\lambda$ = 3 or 5; (b) high water content $\lambda$ = 16 or 20. (c) CN of OW--OW.}\label{fig:09}
\end{figure*}

The structural characteristics of water clusters are further analyzed by studying the RDF and CN between oxygen atoms in water molecules (i.e., OW), as shown in Figure \ref{fig:09}. The first peak of the RDF of OW--OW pairs appears at about 3.45 {\AA}. The peak value is affected by the water content and the ionomer mass content, as shown in Figure \ref{fig:09}(a) and \ref{fig:09}(b), respectively. A cutoff distance of 5.25 {\AA} was used in the calculation of the CN of OW--OW pairs by integrating the RDF, as shown in Figure \ref{fig:09}(c). With the increase in water content and ionomer mass content, the CN of OW--OW increases. With the increase of water molecules, larger water clusters will form. From Figure \ref{fig:09}(a,b), a second peak is observed because of the correlation between different water clusters. With increasing ionomer mass content, the height of the second peak increases and the position slightly shifts, suggesting that the correlation between different water clusters is enhanced. Due to the dense arrangement of the ionomer, the number of charged sites per unit space increases, leading to a longer correlation distance between different water clusters.

\begin{figure*}
  \centering
  \includegraphics[width=1.8\columnwidth]{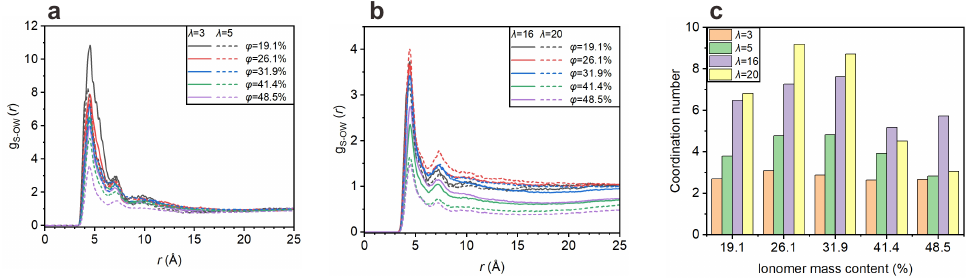}
  \caption{(a,b) RDFs of S--OW pairs at different ionomer mass content at (a) low water contents $\lambda$ = 3 and 5 and (b) high water contents $\lambda$ =16 and 20. (c) CN of S-OW.}\label{fig:10}
\end{figure*}

To reveal the water distribution in CLs, the RDF and corresponding CN of S--OW pairs are analyzed. The RDFs of S--OW pairs are shown in the Figure \ref{fig:10}(a-b). The first peak appears at about 4.45 {\AA}. The RDF was integrated to obtain the CN with a cut-off distance of 6.35 {\AA}. When the ionomer mass content is less than 31.9\%, the height of the first peak is reduced by increasing the water content, but because of the rapid development of water clusters, more water molecules appear in the first shell layer of sulfur. When the ionomer mass content is greater than 41.4\%, the height of the first peak further decreases and the CN also decreases. Because of the dense distribution of the ionomer, the diffusion of the water is inhibited, and the water molecules become cluster form. Hence, the water molecules are more difficult to surround sulfonic acid groups, which is not conducive to proton transport through the charged sites of ionomer side chains.

\begin{figure*}
  \centering
  \includegraphics[width=1.5\columnwidth]{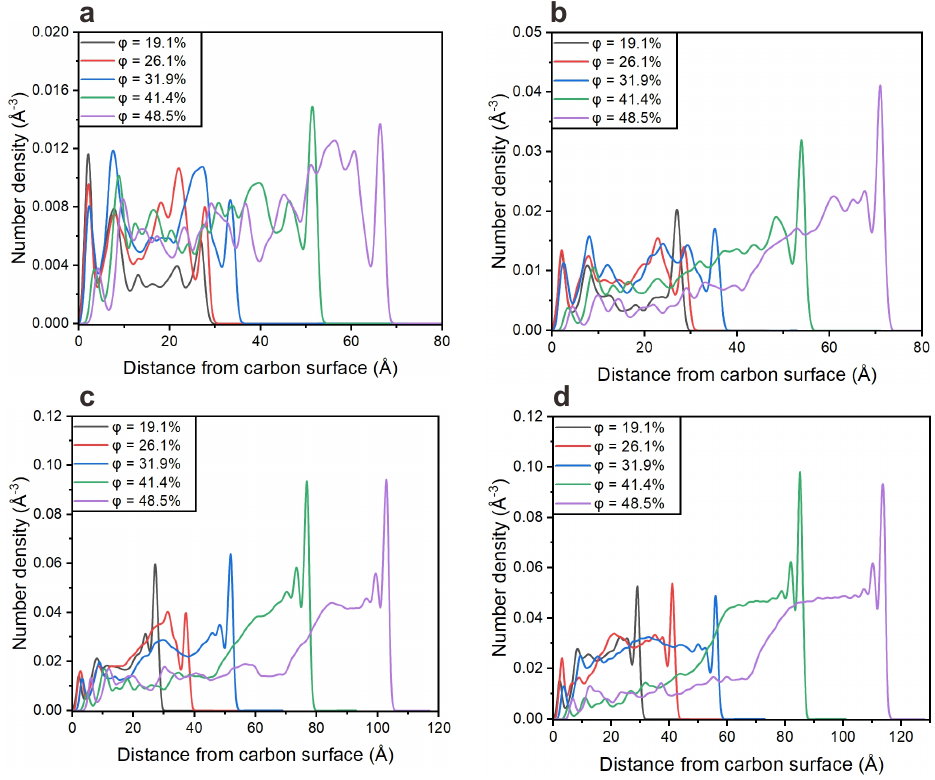}
  \caption{Number density of water molecules on the Pt/C substrate along the thickness direction: (a) $\lambda$ = 3, (b) $\lambda$ = 5, (c) $\lambda$ = 16, and (d) $\lambda$ = 20.}\label{fig:11}
\end{figure*}

To further study the distribution of water molecules, the number density of water is analyzed. As shown in Figure \ref{fig:11}, the distribution of water molecules in the thickness direction (i.e., the $z$ direction) is markedly influenced by the ionomer mass content. When the ionomer mass content is very large (i.e., larger than 41.4\%), the dense ionomer layer will separate a large number of water molecules from the substrate, which form a dense water layer as shown in Figure \ref{fig:11}(a-d). The platinum catalyst is in the range of $0 < z < 30 $ {\AA}. Because of the hydrophilicity of platinum, the number density of water molecules increases with increasing the water content, as shown in Figure \ref{fig:11}(a-d) for $\lambda$ = 3 to 20. When the ionomer mass content is low, this phenomenon is more remarkable. When the ionomer mass content is increased, the number density of water molecules will decrease significantly due to the reduction in free space. In the region further away from the platinum particle ($z > 30$ {\AA}), the number density of water increases along the $z$ direction, and the thickness of the water clusters also increases accordingly due to the increased ionomer mass content, forming a large water cluster extending into the ionomer with multiple slender water channels, providing a continuous channel for proton transport.

\begin{figure}
  \centering
  \includegraphics[scale=0.6]{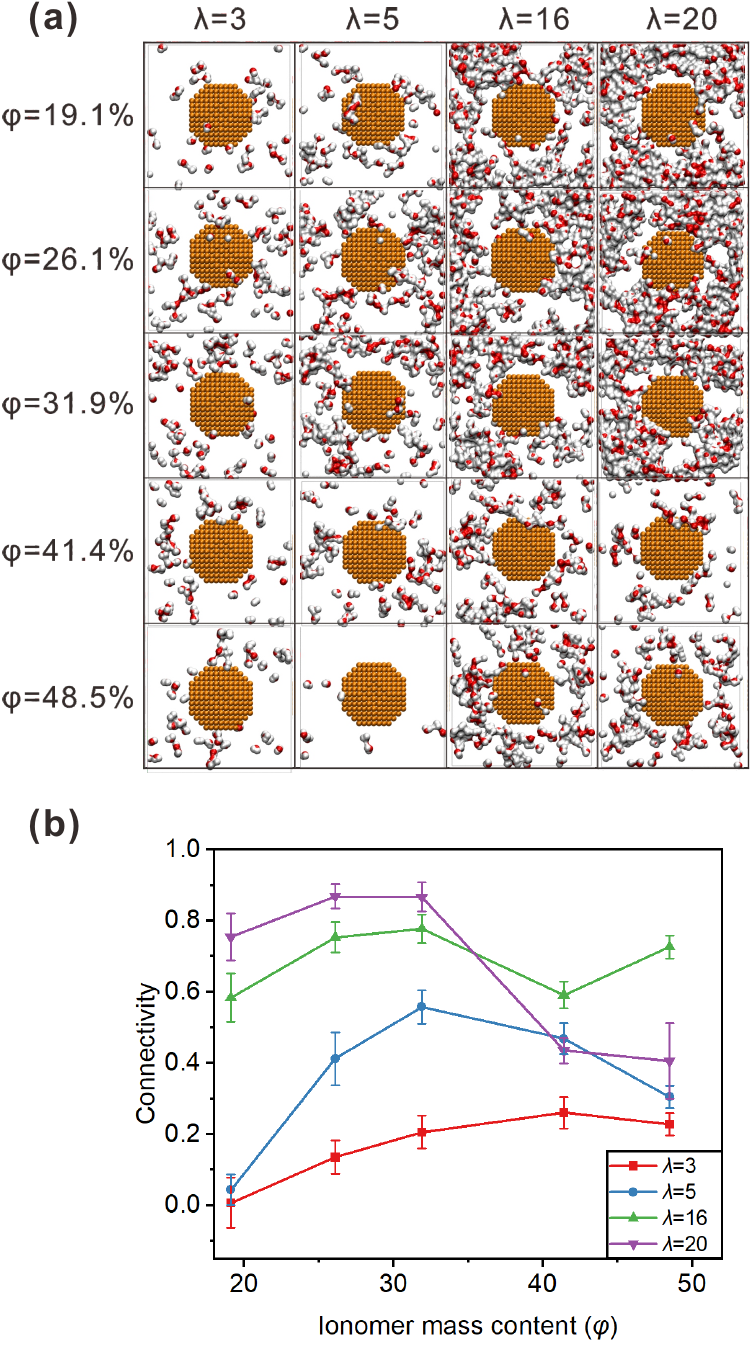}
  \caption{(a) Morphology of water clusters around platinum catalyst at different water contents and ionomer mass contents. (b) Variation of connectivity of water clusters against the water content ($\lambda$) at different ionomer mass contents ($\varphi$).}\label{fig:12}
\end{figure}

To analyze the water molecules around the platinum catalyst, we plot the morphology of the water cluster around the platinum particle, as shown in Figure \ref{fig:12}(a). The connectivity of water clusters is also analyzed to illustrate the morphology of water clusters, as shown in Figure \ref{fig:12}(b). The connectivity   is defined as \cite{mabuchi18}
\begin{equation}\label{eq:04}
  {{C}_{\text{avg}}}=\frac{{{n}_{\text{sulfo}}}-{{n}_{\text{avg}}}}{{{n}_{\text{sulfo}}}-1}
\end{equation}
where ${{n}_{\text{sulfo}}}$ is the number of sulfonic acid groups in the system (from 40--160 in this study) and ${{n}_{\text{avg}}}$ is the average number of water clusters in 100 configurations of the 5-ns NVT simulation. The cluster analysis was performed in OVITO software \cite{stukowski10}. In the calculation of connectivity, it is assumed that each water molecule cluster is formed by relying on the sulfonic acid group \cite{mabuchi18}. In the water cluster analysis, a cut-off distance of 3.5 {\AA} between oxygen atoms in water molecules and hydronium ions is used. When $c = 0$, all water clusters are completely isolated from each other, and when $c = 1$, the water molecules are connected to form a large and continuous water cluster. Hence, the value can reflect the degree of connectivity of water clusters.

From Figure \ref{fig:12}, we can find that the increase in water content has a strong impact on the water cluster connectivity, while the influence of ionomer mass content shows different results under different water contents. At low water contents, water clusters are isolated and have poor connectivity. In this condition, low ionomer mass content makes more free space for water diffusion, and the water clusters are not conducive to interconnection. With increasing the ionomer mass content, the connectivity of water clusters increases at first and then decreases. This is because, with more ionomer, the ionomer will spread compactly to the substrate, resulting in more sulfonic acid groups in the unit space. Hence, the pores inside the ionomer provide diffusion channels for water molecules, which is conducive to the enhancement of water cluster connectivity. When the ionomer mass content is further increased to more than 40\%, the clustering effect is generated, and the ionomers shrink inward, hindering the diffusion of water molecules, thus bringing the water molecules to the state of isolated clusters again. At high water contents, because of the high density of water molecules, the connectivity of water clusters significantly improves, which is conducive to proton transport. However, too-high or too-low polymer mass contents are not conducive to the formation of continuous water channels, and the mechanism is consistent with the circumstances under low water content. Comparing the snapshots of $\lambda$ = 16 and 20 in Figure \ref{fig:12}(a), we can see that at high water contents, the clustering effect of the ionomer will be intensified, affecting the distribution of water clusters and the formation of continuous water channels. Hence, the connectivity between the water clusters will sharply reduce, as shown in Figure \ref{fig:12}(b). The clustering effect will offset the beneficial effects caused by increasing the water molecules, thus it is adversely affecting the transport of protons.

\subsection{Hydronium ion transport properties}\label{sec:3.4}
\begin{figure*}
  \centering
  \includegraphics[width=2.05\columnwidth]{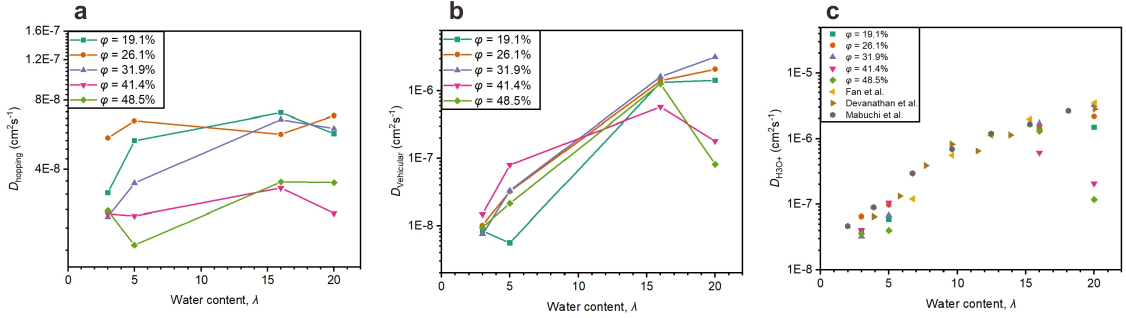}
  \caption{(a) Hopping diffusion coefficient of hydronium ions at different water contents ($\lambda$) and different ionomer mass contents ($\varphi$) and different water contents ($\lambda$). (b) Vehicular diffusion coefficient of hydronium ions at different water contents ($\lambda$) and different ionomer mass contents ($\varphi$). (c) Total diffusion coefficient of hydronium ions at different water contents ($\lambda$) and different ionomer mass contents ($\varphi$). The results from the previous simulation results by Fan et al.\ \cite{fan19}, Devanathan et al.\ \cite{devanathan07}, and Mabuchi et al.\ \cite{mabuchi14} are also shown.}
  \label{fig:13}
\end{figure*}

The transport of hydronium ions is not only affected by water channels, but also by the charged sites of ionomer side chains. Therefore, its transport process is more complex and affected by the structure of water and ionomer.
The proton transport in CCLs mainly consists of two mechanisms, hopping and vehicular mechanisms \cite{kusoglu17, jiao11, agmon1995, Choe08}. The vehicular diffusion coefficient ($D_\text{Vehicular}$) was calculated using Eq.\ (\ref{eq:02}), as shown in Figure \ref{fig:13}(b); and the hopping diffusion coefficient was calculated based on the method developed by Deng and co-workers \cite{deng04, jang05}, which can well describe the hopping diffusion of protons \cite{lawler20}. The hopping diffusion coefficient ($D_\text{Hopping}$) was calculated according to Eq.\ (\ref{eq:dhopping})
\begin{equation}\label{eq:dhopping}
    {{D}_{\text{Hopping}}}=\frac{1}{6Nt}\int_{0}^{t\to \infty }{\sum\limits_{i}^{N}{\sum\limits_{j}^{M}{{{k}_{ij}}r_{ij}^{2}P_{ij}^{{}}\text{d}t}}},
\end{equation}
where $N$ and $M$ are the numbers of hydronium ions and water molecules, $t$ is the hopping diffusion time. $P_{ij}$ is the probability of a proton jumping from hydronium ion $i$ to water molecule $j$, defined as
\begin{equation}\label{eq:pij}
    {{P}_{ij}}=\frac{{{k}_{ij}}}{\sum\limits_{j}^{M}{{{k}_{ij}}}}.
\end{equation}
In addition, $r_{ij}$ is the distance between all proton donors (hydronium ions, $i$) and accepters (water molecules, $j$), which was calculated from the equilibrated trajectories. $k_{ij}$ is the proton transfer rate, calculated by
\begin{equation}\label{eq:kij}
    {{k}_{ij}}(r)=\kappa (T,r)\frac{{{k}_{b}}T}{h}\exp \left( -\frac{{{E}_{ij}}(r)-1/2\text{h}w(r)}{RT} \right)
\end{equation}
where $\kappa (T, r)$ is the tunneling factor \cite{lill01} and $\omega(r)$ is the frequency for zero point energy correction \cite{lill01frAQB}. $k_b$ is the Boltzmann constant and $h$ is the Plank constant. $E_{ij}(r)$ is the energy barrier for proton transport from the donor to the accepter at the distance of $r$. The relative energy change was calculated as a function of the distance between the proton donors and accepters to analyze the energy barrier of proton hopping transport. In addition, the solvent effects were considered by involving the Poisson-Boltzmann self-consistent reaction field models, which were described by Jang and co-workers \cite{jang05}. Finally, we get the $D_\text{H3O+}$ as the total diffusion coefficient of hydronium ions by adding $D_\text{Hopping}$ and $D_\text{Vehicular}$, as shown in Figure \ref{fig:13}(c).

As shown in Figure \ref{fig:13}(c), increasing the water content results in an increase in the diffusion coefficient of hydronium ions for both mechanisms, resulting in an increase in the total diffusion coefficient, which is consistent with previous studies \cite{fan19,devanathan07,mabuchi14}. This is because increasing the water content causes the expansion of water channels and the increase in the connectivity between water clusters. Both the hopping and vehicular mechanisms rely on continuous water channels. Meanwhile, the ionomer mass content also influences the two mechanisms, and the hopping mechanism is more affected than the vehicular mechanism. This is because under the hopping diffusion mechanism, protons are transported by hopping between water molecules, and the charged sites on the ionomer side chain will absorb some protons, which will hinder the occurrence of proton hopping. At low water contents, the diffusion coefficient of hydronium ions is less affected by the ionomer mass content, which is different from the diffusion of water (as discussed in Figure \ref{fig:07}b in Section \ref{sec:3.3}), showing a negative correlation with the mass content of ionomer, see Figure \ref{fig:13}(a, b). Because the proton transport is also affected by the sulfonic acid group, the hopping transport mechanism and the transport through side-chain charged sites play a leading role. This can make up for the decrease of water clusters due to the increased ionomer content, and reduce the influence of the vehicular mechanism. At high water contents, the growth rate of $D_\text{Vehicular}$ is much larger than that of $D_\text{Hopping}$. This is because the region of the ionomer is thickened due to fully hydration, and the transport distance through the side-chain charged sites becomes longer. Therefore, proton transport is dominated by water channels. Under the vehicular transport mechanism, protons will directly transport in water channels in the form of hydronium ions, and the formation of continuous water channels will naturally greatly improve the transport efficiency of the vehicular mechanism. In contrast, the hopping diffusion mechanism will be affected by the more side-chain charged sites due to the thicker ionomer region. Therefore, the transport efficiency of the hopping mechanism increases less. At high ionomer mass contents, due to the clustering effect of the ionomer, the transport of components is hindered in the ionomer layer. The high water content will further intensify the clustering effect, resulting in a significant decrease in the transport efficiency.

\begin{figure*}
  \centering
  \includegraphics[width=1.5\columnwidth]{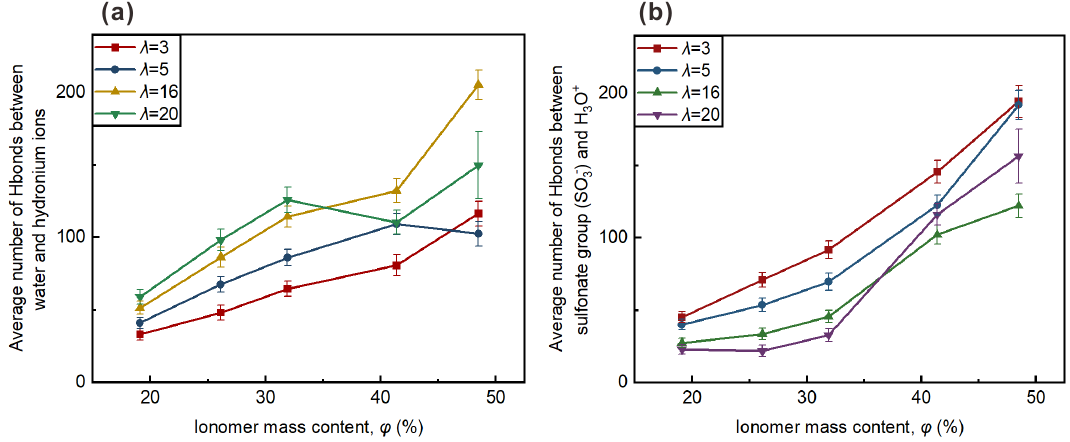}
  \caption{(a) Average number of hydrogen bonds between water and hydronium ions at different ionomer mass contents ($\varphi$) and different water contents ($\lambda$). (b) Average number of hydrogen bonds between sulfonate group ($\text{SO}_3^{-}$) and hydronium ions ($\text{H}_3\text{O}^{+}$) at different ionomer mass contents ($\varphi$) and different water contents ($\lambda$).}
  \label{fig:14a}
\end{figure*}

To further analyze the path of proton transport, we calculated the average number of hydrogen bonds between hydronium ions and water molecules, and between hydronium ions and sulfonic acid groups. The oxygen atoms of hydronium ions were set as donors, and the oxygen atoms of water and sulfonic acid groups were set as accepters. The cut-off distance between donors and accepters was set at 3.5 {\AA}, and the angel cut-off value between donor-hydrogen-accepter was set at 41.7$^\circ$ \cite{kumar07}. The 100 configurations of the last 5-ns sampling process were used for calculation, and the average results are shown in Figure \ref{fig:14a}. As the ionomer mass content increases, the number of hydrogen bonds increases. Because the increase in the ionomer mass content will add more hydronium ions to the system, the total number of hydrogen bonds will naturally increase. Therefore, the influence of the water content on the number of hydrogen bonds is more noteworthy. With the increase in water content, the number of hydrogen bonds between hydronium ions and sulfonic acid groups decreases, while the number of hydrogen bonds between water clusters increases. This further shows that at low water contents, hydronium ions are easy to bind with the charged sites of the ionomer side chains. In this condition, the existence of side chain charged sites plays a key role in the transport of protons. When the water content increases and continuous water channels are formed, the protons will be more inclined to transport in the water channels. At the same time, the clustering effect of high ionomer mass content is also reflected here. At high ionomer mass contents, because of the dense arrangement of ionomer, the number density of sulfonic acid groups increases, and the growth rate of the number of hydrogen bonds between hydronium ions and sulfonic acid groups also significantly increases.

\begin{figure*}
  \centering
  \includegraphics[width=2\columnwidth]{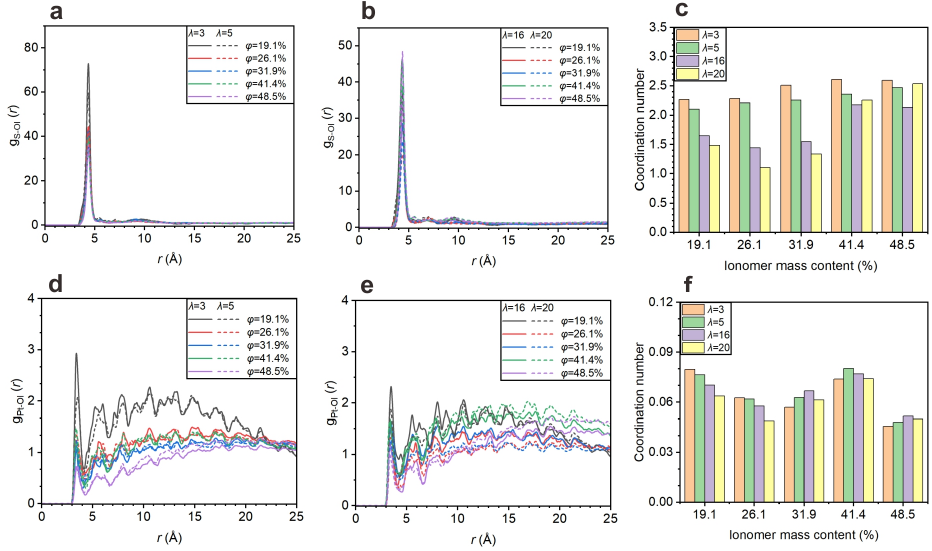}
  \caption{(a,b) RDFs of S--OI pairs at different ionomer mass contents. (c) CN of S--OI. (d,e) RDFs of Pt--OI pairs at different ionomer mass contents. (f) CN of Pt--OI. (a,d) show the results at low water contents, i.e., $\lambda$ = 3 or 5, while (b,e) at high water contents, i.e., $\lambda$ = 16 or 20.}
  \label{fig:15}
\end{figure*}

By analyzing the RDF and coordinate number of the S--OI (here OI refers to oxygen in hydronium ions), the influence of water content and ionomer mass content on the proton transport can be studied. By setting the cutoff distance at 5.05 {\AA}, the coordinate number (CN) of hydronium ions around the sulfonic acid group is calculated by integrating the S--OI RDFs, as shown in Figure \ref{fig:15}(c). It can also be found that the first peaks of S--OI and S--OW are close to each other, indicating that the proton transport relies on water channels, and the height of the first peak is affected by the water content. At the same ionomer mass content, with increasing the water content, the coordinate number of hydronium ions around sulfur decreases, because more water molecules will surround single sulfonic acid groups, which hinders hydronium ions from reaching the sulfonic acid group. With the same water content, as the mass content of the ionomer increases, the ionomer arranges densely which increases the number of sulfonic acid groups (charged sites) per unit space, and then facilitates the adsorption of protons on the charged sites for the proton transport.

By analyzing the distribution of hydronium ions around the catalyst, the effect of water content and ionomer mass content on the efficiency of hydronium ion transport can be obtained. The proton distribution characteristics can be analyzed via the RDF and CN of the Pt--OI. As shown in Figure \ref{fig:15}(d-e), at low water contents, the height of the peak is affected by the ionomer mass content. When the ionomer mass contents increase, it is difficult for hydronium ions to reach the platinum particles because of the isolated distribution of water clusters and the coverage of ionomer on the substrate. When the water content increases, the presence of a large number of water molecules causes the formation of water channel networks. In this condition, proton transport mainly depends on the water channel rather than the ionomer side chain, so it is less affected by the change in the ionomer mass content. As shown in Figure \ref{fig:15}(f), at a low ionomer mass content, the CN of Pt--OI decreases as the water content increases, and when $\varphi$ reaches 48.5\%, the CN decreases sharply. Because the number of charged sites of the ionomer per unit space is small when $\varphi$ is small, the protons are mainly transported through the vehicular mechanism. When the water content increases, the connected water clusters increase the transport distance, resulting in a decrease in the CN. As the number of ionomer molecules increases, the Grotthuss mechanism plays a leading role, the proton transport efficiency is improved, and the CN increases accordingly. When the ionomer mass content reaches 48.5\%, the clustering effect appears, which again hinders the transport of protons.

\section{Conclusions}\label{sec:4}
In this study, all-atom MD simulations are performed to study the transport and distribution of water and protons at different levels of water content and ionomer mass content in CCLs of PEM fuel cells. The RDF and CN of S--S and Pt--S, along with the Connolly surface and density distribution, are analyzed to investigate the morphology and distribution of the ionomer. The carbon skeleton and side chains of the ionomer are attracted by the carbon substrate and platinum catalyst, and ultra-thin dense ionomer layers are formed at the lower region of the platinum and the upper boundary of the system. With increasing the ionomer mass content, the cross-linking effect of S--S is enhanced, which will affect the stacking of ionomer. Then, the well-spread ionomer tends to become clusters and the clustering effect reduces the surface area of the ionomer and hinders proton transport. Volume analysis, diffusion coefficient, along with RDF, CN, and number density analyses are used to study the transport and distribution of water and protons. At low water contents, water molecules form isolated clusters in the pores of the ionomer, and proton transport depends on the charged sites of ionomer side chains. When the ionomer mass content increases, the charged sites per unit space increase, which improves the efficiency of proton transport. After increasing the water content, the isolated water clusters are connected to form continuous water channels, the vehicular mechanism and Grotthuss mechanism play a leading role in proton transport. At high levels of ionomer mass contents, the clustering effect appears, which results in the reduction of space for water diffusion and the accumulation of water above the ionomer. Then, the thickness of the ionomer layer increases, the proton transport path increases, the resistance becomes larger, and the proton transport efficiency decreases. Therefore, the water content and ionomer mass content have an interconnected effect on the distribution and transport of components in fuel cells. By providing physics insight into the mechanism of proton and water transport, the results of this study are helpful for the rational design of catalyst layers of PEM fuel cells.

\begin{acknowledgments}
This work was financially supported by the Department of Science and Technology of Inner Mongolia Autonomous Region (Grant No.\ 2022JBGS0027) and the National Natural Science Foundation of China (Grant Nos.\ 51920105010 and 51921004).
\end{acknowledgments}

\section*{Conflict of Interest Statement}
The authors have no conflicts to disclose.
\section*{DATA AVAILABILITY}
The data that support the findings of this study are available from the corresponding author upon reasonable request.

\section*{REFERENCES}
\bibliography{MDcatalystLayer}
\end{document}